\documentstyle[12pt]{article}
\newcommand{\beq}{\begin{equation}}
\newcommand{\eeq}{\end{equation}}
\newcommand{\bea}{\begin{eqnarray}}
\newcommand{\eea}{\end{eqnarray}}
\newcommand{\rhobar}{\bar{\rho}}

\newcommand{\lkk}{\left[}
\newcommand{\rkk}{\right]}
\newcommand{\lmk}{\left(}
\newcommand{\rmk}{\right)}
\newcommand{\lnk}{\left\{}
\newcommand{\rnk}{\right\}}

\begin{document}

\begin{flushleft}
{\it Yukawa Institute Kyoto}
\end{flushleft}

\begin{flushright}
YITP-96-58 \\
astro-ph/9612014 \\
\end{flushright}
\vspace{1.0cm}

\begin{center}
\Large\bf{On the Jeans instability during \\ 
the QCD phase transition}
\vskip 1.0cm

\large{Michiyasu NAGASAWA}
\footnote{Electronic address: nagasawa@yukawa.kyoto-u.ac.jp}

and

\large{Jun'ichi YOKOYAMA}
\vskip 0.2cm

\large\sl{Yukawa Institute for Theoretical Physics, \\
Kyoto University, Kyoto 606-01, Japan}
\end{center}

\begin{abstract}
The Jeans scale is estimated during the coexistence epoch of
quark-gluon and hadron phases in the first-order QCD phase
transition. It is shown that, contrary to previous claims,
reduction of the sound speed is so little that the phase
transition does not affect evolution of cosmological density
fluctuations appreciably.
\end{abstract}
\newpage

\baselineskip 0.8cm

As is well known, the Jeans length scale, above which density
fluctuations can grow, is given by the sound speed $c_s$ multiplied by 
the dynamical time scale. In the radiation-dominant stage of the
early universe, $c_s$ is equal to $3^{-1/2}$ times the light speed,
which we take to be unity here, so density fluctuations can grow only
on super-horizon scales.

Recently, however, Jedamzik \cite{Jed} and, independently, Schmid,
Schwarz and Widerin \cite{SSW} proposed that if a cosmological
quark-hadron phase transition is of first-order, which is supported
by some of the recent lattice simulations\cite{lattice}, the sound speed
effectively vanishes during the coexistence period of the quark-gluon
and the hadron phases. If this were true, the Jeans scale would become
vanishingly small, and even sub-horizon-scale fluctuations should grow
during this epoch. This temporal decline of the Jeans length
would imprint a nontrivial feature on the
processed spectrum of fluctuations around the solar-mass scale.
\cite{SSW} Furthermore, it has been suggested that such anomalous
growth of perturbations could induce efficient formation of primordial 
black holes (PBH's)\cite{Jed}. The resultant mass function of the 
PBH's would have a peak on the solar-mass scale which is about the
scale of MACHO's.
Thus the idea raised in Refs. 1) and 2) is very interesting
and may be astrophysically important. Unfortunately, however, the
actual evolution of the sound speed, as well as that of the Jeans
length, has not been given in either paper, and their claims are
limited to a qualitative level.

In the present paper we present the result of an explicit calculation of 
the sound speed using the bag model\cite{bag}. We first review how the
quark-hadron phase transition proceeds in this model and then
calculate the sound speed under two distinct conditions.

In the bag model \cite{bag,FMA} the energy density
of the quark-gluon phase, $\rho_q$, is the sum of that
of the relativistic particles and the QCD vacuum energy density, namely,
the bag constant, $B$.
\beq
\rho_q(T) =\frac{\pi^2}{30}g_qT^4+B\ ,
\eeq
where $g_q$ is the effective degree of relativistic freedom.
$g_q\cong 51.25$ when the number of species of relativistic quarks
is two, and $g_q\cong 61.75$ when it is three. The pressure, $p_q$,
and the entropy density, $s_q$, are given by
\bea
p_q(T) &=& \frac{\pi^2}{90}g_qT^4-B\ ,\\
s_q(T) &=& \frac{4\pi^2}{90}g_qT^3\ ,
\eea
respectively.
On the other hand, the energy density, $\rho_h$, the pressure, $p_h$,
and the entropy density, $s_h$, of the hadron phase are
expressed as
\bea
\rho_h(T) &=& \frac{\pi^2}{30}g_hT^4\ ,\\
p_h(T) &=& \frac{\pi^2}{90}g_hT^4\ ,\\
s_h(T) &=& \frac{4\pi^2}{90}g_hT^3\ ,
\eea
where $g_h\cong 17.25$ is the degree of relativistic freedom in
this phase. At the coexistence temperature, $T_c$,
we have $p_q(T_c)=p_h(T_c)$, and hence
it is related to the bag constant by
\beq
B=\frac{\pi^2}{90}\left(g_q-g_h\right)T_c^4. \label{bagcon}
\eeq

In this model, the phase transition from the high-density quark-gluon
phase to the low-density hadron phase starts with bubble nucleation,
whose probability per unit time per unit volume,
$P(T)$, is given by \cite{FMA}
\beq
P\left( T\right) \simeq CT_c^4 
\exp\left(-\frac{16\pi}{3}\frac{\sigma^3}{T_cL^2\eta^2}\right)\ ,
\label{prob}
\eeq
where $L$ is the latent heat per unit volume of the phase transition,
$\sigma$ is the free energy per unit surface area of the bubble, $C$
is a constant, 	and
$\eta$, which represents the degree of super cooling, is defined by
\beq
\eta \equiv \frac{T_c-T}{T_c}\ .
\eeq
The exponent of $P(T)$ is singular at $T_c$, and $P(T)$ vanishes then,
but it increases drastically when $T$ drops slightly from $T_c$. 

Hence the phase transition is expected
to proceed as follows. In the course of
the cosmic evolution the cosmic temperature decreases,
and until $T=T_c$ the entire universe is occupied by
the quark-gluon phase. At this temperature, the hadron phase
does not appear yet, and the super cooling ($T<T_c$) occurs.
But only a small degree of super cooling, $\eta \sim 10^{-3}$,
actually takes place since $P(T)$ soon becomes large\cite{FMA}.
Then the nucleation of the hadron phase starts, and the universe
is reheated by the latent heat up to $T_c$ so that the pressure
equilibrium between two phases is realized and subsequent nucleation
is suppressed. Afterwards, the phase transition proceeds
through expansion of nucleated bubbles whose speed is small
and estimated to be around $ 10^{-3}$\cite{KK},
keeping the balance between heating due to continuous
liberation of latent heat and cooling caused by cosmic expansion.
The temperature and the pressure remain constant in this stage,
but the mean energy density,
\beq
 \rhobar\equiv f_q\rho_q+\left(1-f_q\right)\rho_h\ ,
\eeq
decreases gradually from $\rho_q(T_c)$ to $\rho_h(T_c)$ as the volume
fraction of the quark-gluon phase, $f_q$, drops. During this epoch the
scale factor expands by a factor of about $(g_q/g_h)^{1/3} \simeq 1.4$. 

It is in this coexistence regime that the authors of Refs. 1) and
2) suggested the sound speed vanishes. According to Ref. 1),
an adiabatic compression of the system in the equilibrium two-phase
mixture induces conversion of the low-energy hadron phase into
high-energy quark phase, and then the mean energy density grows due
to the change of $f_q$, with the pressure remaining constant. This
implies that pressure response to a change in energy density is
anomalously small, and hence, by their argument, the sound speed
should effectively vanish.

In order to clarify whether such an intuitive discussion is correct,
we must first find a formula for the sound speed in an inhomogeneous
medium consisting of the mixture of two different phases. While
such a problem has not been fully discussed in any cosmological
context, it is a rather familiar issue in the field of engineering,
e.g., propagation of sound waves in a water-vapor system in a
boiler etc\cite{Aka}.

In the rest frame of the sound wave front\footnote{
Since the time scale of QCD is much shorter than the cosmic
expansion scale, the following discussion without the effect of cosmic 
expansion applies to our problem unchanged.}, suppose that the medium
ahead of the wave front is characterized by a pressure $p$,
the temperature $T$, energy densities $\rho_q$ and $\rho_h$,
volume fraction of the quark-gluon phase $f_q$, and flow velocity $c_s$,
and that the medium behind it is characterized by $p+dp$, $T+dT$,
$\rho_q+d\rho_q$, $\rho_h+d\rho_h$, $f_q+df_q$, and $c_s+du$,
respectively. Then the continuity equation reads
\bea
& & f_q\lmk p+\rho_q\rmk \gamma^2 c_s+
\lmk 1-f_q\rmk \lmk p+\rho_h\rmk \gamma^2 c_s \nonumber \\
&=&\!\!\!\! \lmk f_q+df_q\rmk \lmk p+\rho_q+dp+d\rho_q\rmk
\tilde{\gamma}^2 \lmk c_s-du\rmk \nonumber \\
&+&\!\!\!\! \lmk 1-f_q-df_q\rmk \lmk p+\rho_h+dp+d\rho_h\rmk
\tilde{\gamma}^2 \lmk c_s-du\rmk \ , \label{cont}
\eea
per unit area, where $\gamma=(1-c_s^2)^{-1/2}$ and
$\tilde{\gamma}=\lnk 1-(c_s-du)^2\rnk ^{-1/2}$ are the Lorentz factor
\footnote{We thank Dr. Schwarz for pointing out that
the nonrelativistic equations were employed in the original version
of our paper.}.
The momentum equation is given by
\bea
& & p+f_q\lmk p+\rho_q\rmk \gamma^2 c_s^2+\lmk 1-f_q\rmk \lmk p
+\rho_h \rmk \gamma^2 c_s^2 \nonumber \\
&=&\!\!\!\! p+dp+\lmk f_q+df_q\rmk \lmk p+\rho_q
+dp+d\rho_q\rmk \tilde{\gamma}^2 \lmk c_s-du\rmk ^2 \nonumber \\
&+&\!\!\!\! \lmk 1-f_q-df_q\rmk \lmk p+\rho_h+dp+d\rho_h\rmk
\tilde{\gamma}^2 \lmk c_s-du\rmk ^2 . \label{momentum}
\eea
From (\ref{cont}) we find
\beq
du =c_s\frac{1-c_s^2}{1+c_s^2}
\frac{f_q\lmk dp+\rho_q\rmk +(1-f_q)\lmk dp+d\rho_h
\rmk +\lmk\rho_q-\rho_h\rmk df_q}
{f_q\lmk p+\rho_q \rmk +\lmk 1-f_q\rmk \lmk p+\rho_h \rmk}
=c_s\frac{1-c_s^2}{1+c_s^2}\frac{dp+d\rhobar}{p+\rhobar}\ . \label{du}
\eeq
Inserting this into (\ref{momentum}), we obtain 
\beq
  c_s=\lmk\frac{d\rhobar}{dp}\rmk^{-\frac{1}{2}}\ . \label{onsoku}
\eeq
Thus the sound speed in a mixture is obtained by simply replacing
the energy density $\rho$ with the mean value
$\rhobar$ in the usual formula for a single-component system,
\beq
  c_s=\lmk\frac{dp}{d\rho}\rmk^{\frac{1}{2}}.
\eeq
The pressure derivative of $\bar{\rho}$ is written by
\beq
\frac{d\bar{\rho}}{dp}=f_q\frac{d\rho_q}{dp}
+\left(1-f_q\right)\frac{d\rho_h}{dp}
+\left(\rho_q-\rho_h\right)\frac{df_q}{dp}\ .\label{drhodp}
\eeq
When the sound speed in the hadron phase and the quark-gluon phase
equals to that in the pure radiation, the sum of the first and second
terms of the equation (\ref{drhodp}) provides the value of $3^{-1/2}$.
Hence the difference from the single-component system is given by
the third term and since $\left(\rho_q-\rho_h\right)$ is positive,
if the volume ratio of the quark phase increase as the pressure
becomes higher, the sound speed in the complex phase should be
reduced. In order to evaluate the degree of reduction,
the estimation of $df_q/dp$ is necessary.

The next question is under what conditions we should calculate the
derivative (\ref{onsoku}). Usually one evaluates the sound speed under 
adiabatic conditions, because a change of the entropy associated with
sound-wave propagation is small in most cases of interest. Since this
is also the condition Jedamzik proposed to adopt\cite{Jed}, let us
first calculate the sound speed under this condition. 

The isentropic condition reads
\bea
f_qs_q \gamma c_s+\left(1-f_q\right)s_h \gamma c_s 
 &=& \left(f_q+df_q\right)\left(s_q+ds_q\right)
\tilde{\gamma} \left(c_s-du\right) \nonumber \\
&~&+\left(1-f_q-df_q\right)
\left(s_h+ds_h\right)\tilde{\gamma} \left(c_s-du\right)\ .
\eea
Then using the average entropy density, $\bar{s}$, defined by
\beq
\bar{s}=f_qs_q+\left(1-f_q\right)s_h\ ,
\eeq
and combining this with the continuity equation (\ref{cont}), we find
\beq
\frac{d\bar{\rho}}{p+\bar{\rho}}=\frac{d\bar{s}}{\bar{s}}\ .\label{rhos}
\eeq
Using the relation $\bar{\rho}=\bar{s}T-p$, we can see
this is nothing but the second law of thermodynamics which gives
us no information about $df_q/dp$. The adiabatic condition is
useless for the purpose of calculating the sound speed qualitatively.
Hence we must find a more appropriate one, understanding the
dynamics of the phase transition better.

What is often adopted in the literature \cite{Aka} 
apart from the isentropic condition is the conservation of the
{\it quality} parameter, $x$, which is the energy fraction of
the gas component. The theoretical sound speed in a water-vapor
system evaluated under this condition agrees with experimental
values qualitatively\cite{Aka}. In the present case the corresponding
quantity may be defined by the energy fraction of the high-energy
quark-gluon phase, $x_q$, as
\beq
x_q=\frac{f_q\rho_q}{f_q\rho_q+\left(1-f_q\right)\rho_h}\ ,
\eeq
with which $df_q/dp$ is expressed as
\beq
\frac{df_q}{dp}=\frac{f_q\left(1-f_q\right)}{\rho_h}\frac{d\rho_h}{dp}
-\frac{f_q\left(1-f_q\right)}{\rho_q}\frac{d\rho_q}{dp}
-\frac{f_q\left(1-f_q\right)}{x_q\left(1-x_q\right)}\frac{dx_q}{dp}\ .
\eeq
Since we cannot measure the sound speed experimentally in our
case unlike in a water-vapor system, we must work out the appropriate
value of $dx_q/dp$ from a theoretical view point alone. Here
we propose to calculate the sound speed under the condition
$dx_q/dp=0$. This is appropriate 
because the transition of the phases through
bubble nucleation is totally suppressed at the coexistence temperature,
as seen in (\ref{prob}), and the expansion speed of bubbles is so small 
that energy transfer through bubble expansion or contraction is also
expected to be negligible during sound-wave propagation.

Therefore we substitute this condition into
(\ref{drhodp}) and find
\beq
\left.\frac{d\bar{\rho}}{dp}\right|_{x_q}= 3\rhobar
\left(\frac{f_q}{\rho_q}+\frac{1-f_q}{\rho_h}\right)\ .
\eeq
We thus obtain the sound speed under the constant-quality condition as
\beq
 c_s=\frac{1}{\sqrt{3}}\lnk\lmk y+\frac{1}{y}-2\rmk
 \lkk - \lmk f_q-\frac{1}{2}\rmk^2 +\frac{1}{4}\rkk
 +1\rnk^{-\frac{1}{2}}\ . \label{result}
\eeq
Here $y$ is defined by
\beq
y\equiv\frac{\rho_q(T_c)}{\rho_h(T_c)}=\frac{4g_q-g_h}{3g_h},  \label{y}
\eeq
where we have used (\ref{bagcon}).
Since $y+1/y \geq 2$, we find from (\ref{result}) that the sound speed 
in the mixed state is indeed smaller than that in the case of pure
radiation and it can be arbitrarily small if $y$ is much larger or much
smaller than unity. In the present case, however, the actual value of
$y$ is given by $y=3.63$ for $g_q=51.25$ and $y=4.44$ for $g_q=61.75$.
Hence reduction of the sound speed is mild, and even the minimum
sound speed, which is realized at $f_q=0.5$, is as large as
\beq
c_s^{\rm min}=\left(0.77 \sim 0.82 \right)\times \frac{1}{\sqrt{3}}
\eeq
for $g_q=61.75$ and $51.25$, respectively.

Thus the intuitive claim that $c_s \approx 0$ is proved
to be inappropriate as long as we consider the sound speed
at the coexistence temperature. Although
it is true that the coexistence of different phases
at the first order phase transition reduces the velocity of the sound
wave, its degree in the case of the quark-gluon to hadron QCD phase
transition is insignificant and has no drastic effect on the
development of cosmological density perturbations.

Turing back to the original expression (\ref{drhodp}), we see that
the only way to realize a vanishingly small sound speed is to have
an extremely large value of $df_q/dp$. But this is not possible at
the coexistence temperature $T_c$ since the nucleation rate
(\ref{prob}) vanishes then. Nevertheless it may be possible
to realize a large value of $df_q/dp$ at a smaller temperature
when the nucleation rate rises significantly, and the sound speed
may become vanishingly small then. Unfortunately, however, in the
actual evolution of the universe, in the bag model the degree of
super cooling is expected to be very small, as mentioned above, so
even if the sound speed vanished in this era, the duration of this
era is too short to leave any trace in the spectrum of density
fluctuations.

Until now we have assumed that the QCD phase transition is of first
order and subcritical fluctuations are negligible. If, on the contrary,
it were of second order, and subcritical fluctuations were effective,
the phase transition would proceed while maintaining an equilibrium
population between quark-gluon and hadron phases. Under such a
circumstance, $f_q$ would change only mildly against perturbation
by a sound wave. Thus $df_q/dp$ cannot take a large enough value to
suppress the sound speed drastically in these cases.

We therefore conclude that the quark-hadron phase transition has 
no appreciable effect on the final spectrum of density fluctuations.

\vskip 1cm 
MN acknowledges support from JSPS fellowship.
This work was partially supported by the Japanese Grant
in Aid for Science Research Fund of the Ministry of Education,
Science, Sports
and Culture Nos. 5110(MN) and 08740202(JY).

%\newpage

\end{document}